\documentclass[a4paper]{article}
\usepackage[latin1]{inputenc}
\usepackage[T1]{fontenc}
\usepackage[english,english]{babel}
\usepackage{amsmath}
\usepackage{amssymb,amsfonts,textcomp}
\usepackage{color}
\usepackage{array}
\usepackage{supertabular}
\usepackage{hhline}
\usepackage{hyperref}
\usepackage{graphicx}
\usepackage{caption}
\hypersetup{pdftex, colorlinks=true, linkcolor=blue, citecolor=blue, filecolor=blue, urlcolor=blue, pdftitle=, pdfauthor=, pdfsubject=, pdfkeywords=}
\newcommand\textsubscript[1]{\ensuremath{{}_{\text{#1}}}}
\newcommand\textstyleInternetlink[1]{\foreignlanguage{english}{\textcolor[rgb]{0.0,0.0,0.5019608}{#1}}}
\makeatletter
\newcommand\arraybslash{\let\\\@arraycr}
\makeatother
\makeatletter
\newcommand\ps@Standard{
  \renewcommand\@oddhead{}
  \renewcommand\@evenhead{}
  \renewcommand\@oddfoot{}
  \renewcommand\@evenfoot{}
  \renewcommand\thepage{\arabic{page}}
}
\makeatother
\title{}
\author{}
\date{2013-05-28}
\begin{document}
\clearpage\setcounter{page}{1}\pagestyle{Standard}
{\centering\selectlanguage{english}\bfseries
Augmenting transcriptome assembly combinatorially
\par}

\bigskip

{\centering\selectlanguage{english}
Prachi Jain\textsuperscript{1+}, Neeraja M Krishnan\textsuperscript{1+} and Binay
Panda\textsuperscript{1,2 *}
\par}

\bigskip

{\centering\selectlanguage{english}
\textsuperscript{1}Ganit Labs, Bio-IT Centre, Institute of Bioinformatics and Applied Biotechnology,
Bangalore, India
\par}

{\centering\selectlanguage{english}
\textsuperscript{2}Strand Life Sciences, Hebbal, Bangalore, India
\par}

\bigskip

{\centering\selectlanguage{english}
\textsuperscript{+}equal contribution
\par}

{\centering\selectlanguage{english}
*corresponding author (\href{mailto:binay@ganitlabs.in}{\textstyleInternetlink{binay@ganitlabs.in}})
\par}

\bigskip

{\selectlanguage{english}\bfseries
Abstract}

{\selectlanguage{english}
RNA-seq allows detection and precise quantification of transcripts, provides comprehensive
understanding of exon/intron boundaries, aids discovery of alternatively spliced isoforms and
fusion transcripts along with measurement of allele-specific expression. Researchers interested in
studying and constructing transcriptomes, especially for non-model species, often face the
conundrum of choosing from a number of available \textit{de novo} and genome-guided assemblers. \ A
comprehensive comparative study is required to assess and evaluate their efficiency and sensitivity
for transcript assembly, reconstruction and recovery. None of the popular assembly tools in use
today achieves requisite sensitivity, specificity or recovery of full-length transcripts on its
own. Hence, it is imperative that methods be developed in order to augment assemblies generated
from multiple tools, with minimal compounding of error. Here, we present an approach to
\ combinatorially augment transciptome assembly based on a rigorous comparative study of popular
\textit{de novo }and genome-guided transcriptome assembly tools.}

\bigskip

{\selectlanguage{english}\bfseries
Introduction}

{\selectlanguage{english}
High-throughput technology has changed our understanding of many facets of biology, from diseases
\ to plant genetics , and to synthetic biology . The advent of DNA microarrays in the
90{\textquoteright}s ushered an era of high-throughput genome-wide gene expression profiling
studies . DNA microarray, although a powerful technique, is dependent on gene annotation, and hence
genome sequence information. This is circumvented by RNA sequencing (RNA-seq), which uses
next-generation sequencing instruments. This shift from the semi-quantitative, hybridization-based
approaches, as in DNA microarrays, to the quantitative, sequencing-based approaches has
tremendously facilitated gene expression analysis. RNA-seq experiments yield additional information
on transcriptome characterization and quantification, including strand-specificity, mapping of
fusion transcripts, small RNA identification and alternate splicing information . }

\bigskip

{\selectlanguage{english}
A number of tools have been developed for transcriptome assembly in the past. They are classified
under \textit{de novo }and genome-guided assembly categories. Trinity , SOAPdenovo-Trans , Oases
\ and Trans-ABySS \ are \textit{de novo} tools, while TopHat-Cufflinks, \ and Genome guided Trinity
\ fall under the second category. \textit{De novo} assembly tools create short contigs from
overlapping reads, which are extended based on insert size estimates. They could be based on either
construction of \textit{de Bruijn} graphs (using \textit{k}{}-mers for short reads) or use
overlap-layout-consensus (for longer reads) algorithms . In Genome guided Trinity, the reads are
aligned to the genome and partitioned into read clusters, which are individually assembled using
Trinity. TopHat-Cufflinks is an aligner cum assembly pipeline, involving spliced read alignment to
the genome and subsequent assembly of these aligned reads into transcripts. The genome template and
read pairing are utilized here to produce full-length or near full-length transcripts. Studies have
identified \textit{k-mer} size as an important parameter for short-read transcriptome assembly
quality. The theoretical calculation on assembling transcriptomes based on \textit{k-}mer size (a
lower or higher \textit{k}{}-mer length for low- or high-expressed genes respectively) \ has also
been backed by experimental evidence . The \textit{k}{}-mer length depends on the genome
complexity, sequencing depth and sequencing error . One can take a decision on \textit{k}{}-mer
length depending on desired transcript diversity vs continuity of a given transcriptome. }

\bigskip

{\selectlanguage{english}
Varying expression levels among genes, presence of homologues and spliced isoforms increase the
complexity in transcriptome assembly. Availability of genome sequence can aid the performance of
transcriptome assembly process. In its absence, genomes of closely related organisms can be used
for transcriptome assemblies . Both \ \textit{de novo} and genome-guided approaches have their own
strengths and one can think of combining data from both to achieve higher sensitivity. However, the
process of combining assemblies from multiple assemblers is not error free and may produce false
assemblies. In the current paper, we first describe a detailed analysis of existing \textit{de
novo} and genome-guided tools and then determine the best combination of tools that can be used to
augment the process of assembly while keeping the false assemblies to a minimum. }

\bigskip

{\selectlanguage{english}\bfseries
Materials and Methods}

\bigskip

{\selectlanguage{english}\itshape
Simulating RNA-seq reads}

{\selectlanguage{english}
The \textit{Arabidopsis thaliana} (TAIR10) complete genome sequence, co-ordinates for genes and
transposons were downloaded from
\url{ftp://ftp.arabidopsis.org/home/tair/Genes/}\href{ftp://ftp.arabidopsis.org/home/tair/Genes/TAIR10_genome_release}{\textstyleInternetlink{TAIR10\_genome\_release}}
and the GFF file was parsed to obtain exonic coordinates. These were used to simulate Illumina-like
RNA-seq reads using Flux-simulator \ (FS-nightly-build\_1.1.1-20121119021549) with the following
options supplied within a parameter file (Supplementary Methods): NB\_MOLECULES 5000000 ,
SIZE\_DISTRIBUTION N(300,30) , READ\_NUMBER 4000000 , READ\_LENGTH 76 , PAIRED\_END YES , ERR\_FILE
76. The resultant Illumina-like reads were split into 2 fastq files corresponding to read1 and
read2 using a Python script from the Galaxy tool suite (Supplementary Methods).}

\bigskip

{\selectlanguage{english}\itshape
Zebrafish transcriptome data}

{\selectlanguage{english}
RNA-seq reads for the zebrafish,\textit{ Danio rerio,} (2dpf) embryo were downloaded from SRA
(ERR003998) . The zebrafish genome (Zv9) downloaded from Ensemble was used for genome-guided
transcriptome assembly. The transcripts for hox gene cluster (listed in Table 2 in Corredor-Adamez
et al. 2005) were downloaded from ENSEMBL, and used to identify hox-related genes from the
assemblies through Megablast. We focused on read-covered shared and unique regions of the hox gene
cluster. }

\bigskip

{\selectlanguage{english}\itshape
Read assembly }

{\selectlanguage{english}
The simulated and \textit{Danio rerio} reads were assembled using four \textit{de novo} (Trinity,
version r2012-06-08; Trans-ABySS, version 1.3.2; Oases, version 0.2.08; SOAPdenovo-Trans, version
1.0 for simulated data and version 1.02 for \textit{Danio rerio}) and two genome-guided (Tophat1,
version 2.0.4 that uses Bowtie1 , version 0.12.8 and Cufflinks, version 2.0.0; Genome guided
Trinity, version r2012-10-05 that uses GSNAP , version 2012-07-20 [V3] for simulated data and
version 2013-03-31 [V5]) for zebrafish data in the process of transcriptome assembly pipelines. We
used Trinity and Genome guided Trinity with a fixed kmer size of 25bp, Trans-ABySS on ABySS
\ (version 1.3.3) multi-kmer assemblies (ranging from 20-64bp), Oases on Velvet \ (version 1.2.07)
multi-kmer assemblies (every alternate kmer ranging from 19-71bp) and SOAPdenovo-Trans with a fixed
kmer size of 23bp. We tested the Tophat2 \ (version 2.0.7 that uses Bowtie version 2.0.5)-Cufflinks
pipeline on the simulated data but did not observe any difference in the assembly statistics
compared to the Tophat1-Cufflinks pipeline that uses Bowtie1. All assemblers were run using default
parameters (details in Supplementary Methods). However, we fixed the parameter for minimum length
of assembled fragment as 76bp (equal to the length of the read). In the case of SOAPdenovo-Trans,
contigs were used instead of scaffolds for all downstream analyses, as the minimum length cutoff
could not be set for scaffolds. }

{\selectlanguage{english}
In order to test whether we could augment the assembly of Trinity-predicted transcript fragments
with those from Tophat1-Cufflinks, we used Megablast to map the Trinity transcript fragments
against the Tophat1-Cufflinks transcripts, and then augmented Trinity assembly with those
transcript fragments unique to Tophat1-Cufflinks assembly (see Supplementary Methods for details).
}

\bigskip

{\selectlanguage{english}\itshape
Redundancy removal using CD-HIT-EST}

{\selectlanguage{english}
We used CD-HIT-EST \ (version 4.5.4) to remove redundancy in each assembly. It retained the longest
sequence out of a cluster of sequences that shared at least 95\% sequence similarity based on a
word size of 8. Further details on the options used to run CD-HIT-EST are provided in Supplementary
Methods. }

\bigskip

{\selectlanguage{english}\itshape
Model Assembly}

{\selectlanguage{english}
We defined all read-covered transcript regions (TAIR10 for simulated data and hox gene cluster for
\textit{Danio rerio}) as Model Assembly or MA as previously described in the report by Mundry et al
2012 .}

\bigskip

{\selectlanguage{english}
\textit{Calculation of N50 and N}\textit{\textsubscript{(MA)}}\textit{50 statistics for an
assembly}}

{\selectlanguage{english}
N50 is defined as the minimum contig length for 50\% of the assembly, after sorting the contigs in
the descending order of their lengths. We calculated N50 values for all assemblies, MA and the
TAIR10 simulated transcripts, pre- and post-CD-HIT-EST. We also calculated the N50 values for each
assembler, while taking the MA cumulative size (bp) as the denominator instead of the respective
assembly sizes. We termed these N50 values as the N\textsubscript{(MA)}50. }

\bigskip

{\selectlanguage{english}\itshape
Mapping assemblies to MA using Megablast}

{\selectlanguage{english}
The assembled fragments (query) were mapped against the MA fragments (subject) using Megablast
\ (blast+ version 2.2.26) with default parameters (Supplementary Methods). The Megablast hits were
parsed in order to either maximize query coverage (to compute mis-assembly statistics) or maximize
subject coverage (to compute MA recovery statistics). This was done to choose the best hits and
discard the partially overlapping hits when the unique coverage was lower than or equal to 10bp.
The scripts used to remove nested and partial overlaps from Megablast hits are provided in
Supplementary Methods.}

\bigskip

{\selectlanguage{english}\itshape
Expression level bin categories}

{\selectlanguage{english}
We estimated the average {\textquotedblleft}per nucleotide coverage{\textquotedblright} (pnc) for
all MA fragments, based on their read support as Number of reads multiplied by Read Length divided
by MA fragment length. The MA fragments were then categorized into 8 expression level bins, B1 to
B8, the pnc for each being: 1 for B1, {\textgreater}1 \& {\textless}= 2 for B2, {\textgreater}2 \&
{\textless}= 3 for B3, {\textgreater}3 \& {\textless}= 4 for B4, {\textgreater}4 \& {\textless}= 5
for B5, {\textgreater}5 \& {\textless}=10 for B6, {\textgreater}10 \& {\textless}= 30 for B7, and
{\textgreater}30 for B8. We chose denser sampling for the lower pnc values and sparser sampling for
the higher pnc values since we observed the distribution of MA fragments to be denser in the lower
pnc categories. }

\bigskip

{\selectlanguage{english}\itshape
Recovery of isoforms}

{\selectlanguage{english}
Using simulated data, we obtained the MA equivalent for the exonic regions of isoform-bearing genes,
and performed a Megablast search of the assemblies against it (Supplementary Methods). The
Megablast hits were parsed to maximize subject coverage, after removing nested and partial overlaps
(same as described earlier). For all assemblers, we calculated the number of exons recovered per
isoform and the length recovery of each exon.}

\bigskip

\bigskip

{\selectlanguage{english}\bfseries
Results}

{\selectlanguage{english}
We compared the performance of assemblers using a variety of assembly-based parameters (numbers and
lengths of assembled fragments, N50, N\textsubscript{(MA)}50 and extent of redundancy) and mapping
parameters (mapping based recovery of MA fragments (numbers and lengths), mis-assembly, reliance on
pnc) for isoforms and non-isoforms, and shared and unique transcript regions.}

\bigskip

{\selectlanguage{english}\bfseries\itshape
Assembly statistics}

{\selectlanguage{english}
We simulated reads for the exonic regions of the \textit{Arabidopsis thaliana} TAIR10 genome (see
Methods for details) and obtained Illumina-like paired-end 76bp reads covering 15,532 transcripts.
The transcript regions contiguously covered by reads were termed as Model Assembly (MA) fragments
and were used as a valid reference for mapping the assemblies. A given transcript, therefore,
comprised of one or more MA fragments, the shortest being 76 bp in length. We obtained 70,382 MA
fragments from the 15,532 TAIR10 transcripts. The reads were assembled using Trinity, Trans-ABySS,
Oases, SOAPdenovo-Trans, Tophat1-Cufflinks and Genome guided Trinity (see Methods for details) with
default parameters and a minimum of 76bp assembled fragment length. We compared the numbers of
assembled fragments, the minimum and maximum fragment lengths, their respective length frequency
distributions, the N50 and N\textsubscript{(MA)}50 statistics for the six assemblers, before and
after eliminating redundancy in the assembly using CD-HIT-EST. }

{\selectlanguage{english}
The total number of assembled fragments varied widely across the six assemblers (Table 1). We
observed a 4- and \~{}1.5-fold reduction in assembly size post redundancy removal for Trans-ABySS
and Oases, respectively (Table 1). The shortest fragment reported by all assemblers was 76bp (as
fixed by the minimum reporting length threshhold). The range of fragments at the long end varied
across assemblers with the longest for Tophat1-Cufflinks at 10,502 bp (same as the maximum MA
fragment length, Table 1). This was expected since Tophat1-Cufflinks is a genome-guided assembler
that allows recovery of full length or near full length transcripts. For Trinity, genome guided
Trinity and SOAPdenovo-Trans, we observed no difference pre- and post-redundancy removal across the
entire distribution of assembled fragment lengths (Figure 1). However, Trans-ABySS resulted in a
higher number of longer redundant assembled fragments (Figure 1) and interestingly a low number of
shorter assembled fragments. Like the results from the frequency distribution statistics for long-
and short-assembled fragments, Trinity, genome guided Trinity and SOAPdenovo-Trans yielded N50
values closer to that of MA. However, Trans-ABySS, Oases and TopHat1-Cufflinks yielded N50 numbers
that were much higher than that of MA. After eliminating redundancy, the N50 values were not
affected for most assemblers except for Trans-ABySS (Table 1). In contrast, the
N\textsubscript{(MA)}50 values, which are not dependent on the size of any assembly, were lower
than that of the MA, for Trinity, Genome guided Trinity and SOAPdenovo-Trans, both before and after
redundancy removal. For Trans-ABySS and Oases, the N\textsubscript{(MA)}50 values were relatively
higher, and were reduced \~{}3 and \~{}1.5-fold respectively, after redundancy removal. The
N\textsubscript{(MA)}50 values were higher than that of the MA for Tophat1-Cufflinks, both before
and after redundancy removal. }

\bigskip

\bigskip

\begin{figure}[htbp]
\begin{flushleft}
\includegraphics[scale=1]{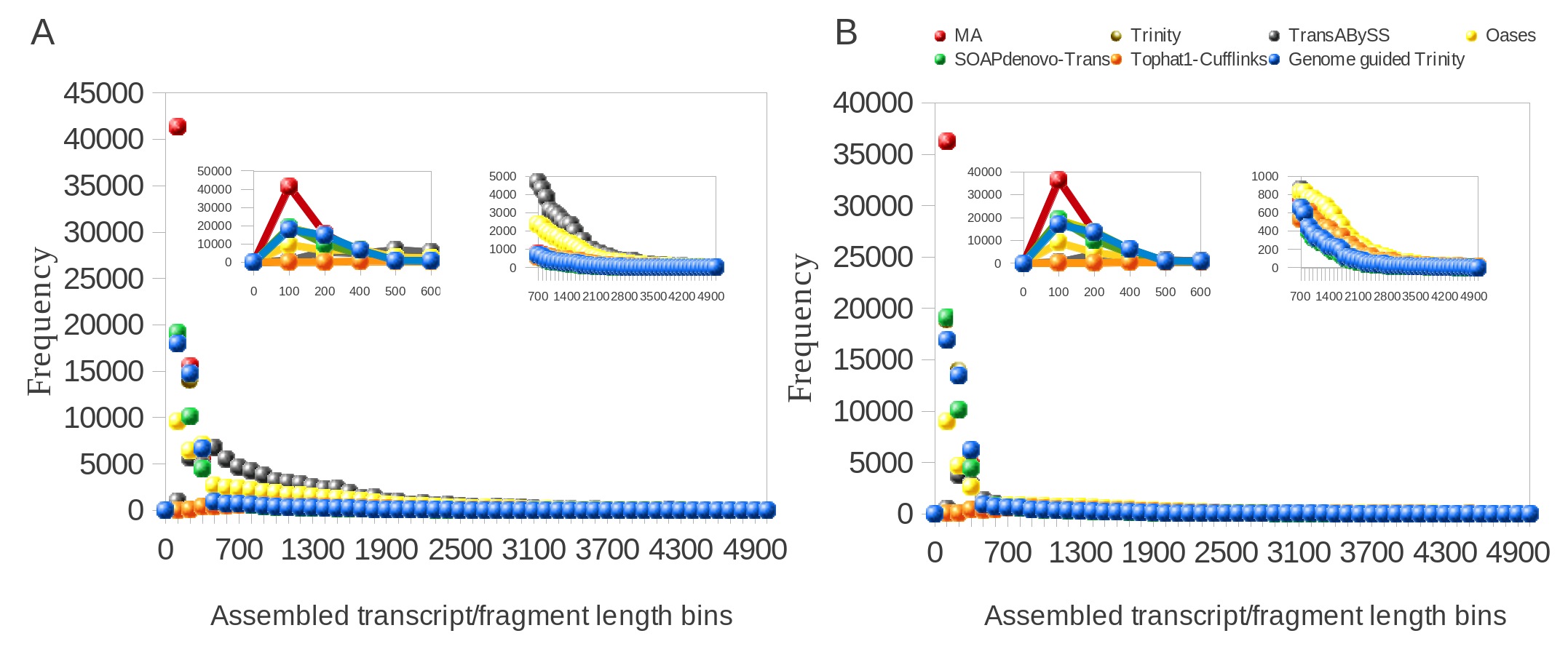}
\caption{Frequency distribution of lengths (nt) of MA \& assembled transcript fragments before (A) and after (B) CDHIT-EST.}
\end{flushleft}
\end{figure}

\begin{figure}[htbp]
\begin{flushleft}
\includegraphics[scale=1.1]{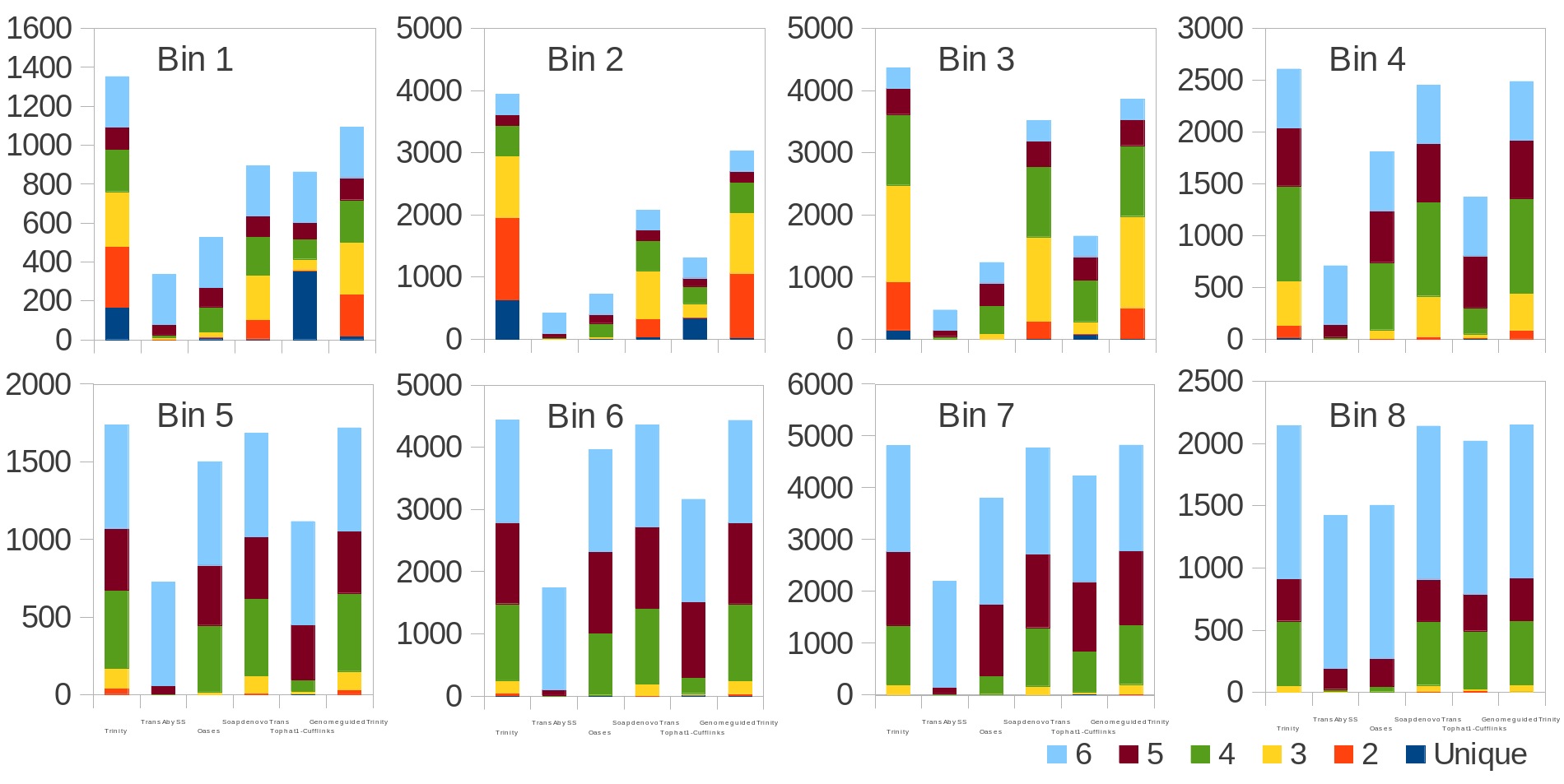}
\caption{The number of recovered MA fragments in different expression bins.}
\end{flushleft}
\end{figure}

\clearpage{\selectlanguage{english}
\foreignlanguage{english}{\textbf{Table 1: }}\foreignlanguage{english}{\textbf{Assembly statistics
pre- and post-CDHit-EST.}}}

\bigskip

{\selectlanguage{english}\bfseries
Pre CDHit-EST}

\begin{center}
\tablehead{}
\begin{supertabular}{m{1.2809598in}m{0.54345983in}m{0.6594598in}m{0.5698598in}m{0.57405984in}m{0.5393598in}m{0.53235984in}m{0.65525985in}m{0.59275985in}}
~
 &
\selectlanguage{english}\bfseries TAIR10 \ transcripts &
\selectlanguage{english}\bfseries Model Assembly (MA) &
\selectlanguage{english}\bfseries Trinity &
\selectlanguage{english}\bfseries Trans-ABySS &
\selectlanguage{english}\bfseries Oases &
\selectlanguage{english}\bfseries SOAP\newline
denovo-Trans &
\selectlanguage{english}\bfseries Tophat1-Cufflinks &
\selectlanguage{english}\bfseries Genome guided Trinity\\
\selectlanguage{english} Total transcripts or fragments &
\selectlanguage{english} 15532 &
\selectlanguage{english} 70382 &
\selectlanguage{english} 45978 &
\selectlanguage{english} 65594 &
\selectlanguage{english} 52533 &
\selectlanguage{english} 39533 &
\selectlanguage{english} 8671 &
\selectlanguage{english} 46064\\
\selectlanguage{english} Transcriptome size (bp) &
\selectlanguage{english} 2.61E+07 &
\selectlanguage{english} 1.67E+07 &
\selectlanguage{english} 1.27E+07 &
\selectlanguage{english} 6.62E+07 &
\selectlanguage{english} 4.21E+07 &
\selectlanguage{english} 1.02E+07 &
\selectlanguage{english} 1.15E+07 &
\selectlanguage{english} 1.27E+07\\
\selectlanguage{english} N50 (bp) &
\selectlanguage{english} 1952 &
\selectlanguage{english} 720 &
\selectlanguage{english} 684 &
\selectlanguage{english} 1390 &
\selectlanguage{english} 1453 &
\selectlanguage{english} 617 &
\selectlanguage{english} 1624 &
\selectlanguage{english} 646\\
\selectlanguage{english} N(MA)50 (bp) &
\selectlanguage{english} 1952 &
\selectlanguage{english} 720 &
\selectlanguage{english} 264 &
\selectlanguage{english} 2802 &
\selectlanguage{english} 2519 &
\selectlanguage{english} 114 &
\selectlanguage{english} 1136 &
\selectlanguage{english} 254\\
\selectlanguage{english} Median transcript/fragment length (bp) &
\selectlanguage{english} 1456 &
\selectlanguage{english} 85 &
\selectlanguage{english} 115 &
\selectlanguage{english} 804 &
\selectlanguage{english} 523 &
\selectlanguage{english} 103 &
\selectlanguage{english} 1124 &
\selectlanguage{english} 121\\
\selectlanguage{english} Min. transcript/fragment length (bp) &
\selectlanguage{english} 201 &
\selectlanguage{english} 76 &
\selectlanguage{english} 76 &
\selectlanguage{english} 76 &
\selectlanguage{english} 76 &
\selectlanguage{english} 76 &
\selectlanguage{english} 76 &
\selectlanguage{english} 76\\
{\selectlanguage{english} Max. transcript/fragment length (bp)}

~

\selectlanguage{english}\bfseries Post-CDHit-EST &
\selectlanguage{english} 14623 &
\selectlanguage{english} 10501 &
\selectlanguage{english} 6888 &
\selectlanguage{english} 7326 &
\selectlanguage{english} 8128 &
\selectlanguage{english} 6300 &
\selectlanguage{english} 10502 &
\selectlanguage{english} 7833\\
~
 &
\selectlanguage{english}\bfseries TAIR10 \ transcripts &
\selectlanguage{english}\bfseries Model Assembly (MA) &
\selectlanguage{english}\bfseries Trinity &
\selectlanguage{english}\bfseries Trans-ABySS &
\selectlanguage{english}\bfseries Oases &
\selectlanguage{english}\bfseries SOAP\newline
denovo-Trans &
\selectlanguage{english}\bfseries Tophat1-Cufflinks &
\selectlanguage{english}\bfseries Genome guided Trinity\\
\selectlanguage{english} Total transcripts or fragments &
\selectlanguage{english} 14615 &
\selectlanguage{english} 62528 &
\selectlanguage{english} 45281 &
\selectlanguage{english} 16060 &
\selectlanguage{english} 28051 &
\selectlanguage{english} 39471 &
\selectlanguage{english} 8395 &
\selectlanguage{english} 42968\\
\selectlanguage{english} Transcriptome size (bp) &
\selectlanguage{english} 2.45E+07 &
\selectlanguage{english} 1.56E+07 &
\selectlanguage{english} 1.21E+07 &
\selectlanguage{english} 1.15E+07 &
\selectlanguage{english} 1.85E+07 &
\selectlanguage{english} 1.02E+07 &
\selectlanguage{english} 1.11E+07 &
\selectlanguage{english} 1.20E+07\\
\selectlanguage{english} N50 (bp) &
\selectlanguage{english} 1951 &
\selectlanguage{english} 788 &
\selectlanguage{english} 627 &
\selectlanguage{english} 1194 &
\selectlanguage{english} 1485 &
\selectlanguage{english} 617 &
\selectlanguage{english} 1621 &
\selectlanguage{english} 661\\
\selectlanguage{english} N(MA)50 (bp) &
\selectlanguage{english} 1951 &
\selectlanguage{english} 788 &
\selectlanguage{english} 260 &
\selectlanguage{english} 798 &
\selectlanguage{english} 1680 &
\selectlanguage{english} 155 &
\selectlanguage{english} 1197 &
\selectlanguage{english} 273\\
\selectlanguage{english} Median transcript/fragment length (bp) &
\selectlanguage{english} 1451 &
\selectlanguage{english} 87 &
\selectlanguage{english} 113 &
\selectlanguage{english} 477 &
\selectlanguage{english} 216 &
\selectlanguage{english} 103 &
\selectlanguage{english} 1121 &
\selectlanguage{english} 120\\
\selectlanguage{english} Min. transcript/fragment length (bp) &
\selectlanguage{english} 201 &
\selectlanguage{english} 76 &
\selectlanguage{english} 76 &
\selectlanguage{english} 76 &
\selectlanguage{english} 76 &
\selectlanguage{english} 76 &
\selectlanguage{english} 76 &
\selectlanguage{english} 76\\
\selectlanguage{english} Max. transcript/fragment length (bp) &
\selectlanguage{english} 14623 &
\selectlanguage{english} 10501 &
\selectlanguage{english} 6888 &
\selectlanguage{english} 7326 &
\selectlanguage{english} 8128 &
\selectlanguage{english} 6300 &
\selectlanguage{english} 10502 &
\selectlanguage{english} 7833\\
\end{supertabular}
\end{center}

\bigskip

{\selectlanguage{english}\bfseries\itshape
Mapping-based statistics}

{\selectlanguage{english}
MA fragments from TAIR10 transcripts corresponding to transposable elements and transcript isoforms,
as per the GFF annotation, were not included in the expression level based binning. We excluded
these elements since they have stretches of identical sequences, which posed a problem in assigning
the pnc index and reliable mapping to the correct isoform and/or transposable element. Out of a
total of 70,382 MA fragments, we were thus left with 46,805 fragments, which still had certain
level of sub-sequence similarity, presumably arising from gene paralogs and SSRs. The numbers of MA
fragments obtained in different expression bins were, 15,706 in B1 (the lowest expression bin),
9,722 in B2, 5,205 in B3, 2,797 in B4, 1,820 in B5, 4,522 in B6, 4,862 in B7 and 2,171 in B8 (the
highest expression bin). \ }

{\selectlanguage{english}\itshape
MA recovery statistics}

{\selectlanguage{english}
The assembled transcript fragments from all assemblers were mapped to the MA using Megablast to
assign common identifiers for comparisons. We compared the numbers and lengths of MA fragments
recovered within expression bins, B1 to B8, using the best hits from Megablast. Trinity detected
the highest number of MA fragments in B1 followed by Genome guided Trinity and SOAPdenovo-Trans
(Figure 2). However, the number of unique transcripts detected was highest for Tophat1-Cufflinks,
followed by Trinity for the lowest expression bin B1 (Figure 2). The extent of overlapping between
Trinity and either Genome guided Trinity or SOAPdenovo-Trans was 30\% for the lowest expression bin
B1. The numbers detected were highest for Trinity for all expression bins (B1-B8) with other
assemblers showing recovery with increase in expression. In comparison to Trinity, for Oases and
Trans-ABySS the detection sensitity increased with increased expression but the overall number of
recovered MA fragments remained lower in all expression bins (Figure 2). Tophat1-Cufflinks showed a
drop in its detection sensitivity from B1 to B2, and a steady rise thereafter. For all other
assemblers, with the exception of Trans-ABySS, we observed an increase in higher-order
intersections proportional to the decrease in unique detections and lower order intersections
across bins B2-B8 (Figure 2).}

\bigskip

{\selectlanguage{english}
We observed that longer MA fragments tend to have higher read support, thus assigning them to the
\ higher expression level bins (Supplementary Figure 1). The shorter MA fragments, however, got
assigned to all expression level bins, as they had low-to-high read support (Supplementary Figure
1). Therefore, we decided to compare the length recovery across assemblers, for each expression
level bin, in different MA fragment length categories (Figure 3). For the shortest length category,
76bp, the median length recovery was close to 100\%, across all expression bins for all assemblers,
with the exception of Trans-ABySS (Figure 3). The outlier trend was highly variable across
assemblers, and clustered closer to the median for SOAPdenovo-Trans, followed by Trinity and Genome
guided Trinity. For Trinity, Trans-ABySS, Oases and Genome guided Trinity, we observed the outliers
clustering around \~{}40\% length recovery (\~{}30bp) for the MA fragments in the lowest expression
bin that matched the word size for Megablast (28bp). The outliers in the lowest expression bin for
Tophat1-Cufflinks and those in the highest expression bins for Oases spanned all the way from
\~{}40\% to the median. For the subsequent length categories, we observed a gradual increase in the
median length recovery from as low as 20\% to all the way upto 100\%, with an increase in the
expression levels. This gradual increase was also seen, to a minor extent, for the same bin across
MA length categories, for most assemblers. Tophat1-Cufflinks showed a median recovery of 100\%
across all expression bins and all MA fragment length categories. The pattern of length recovery
was similar for Trinity, SOAPdenovo-Trans and Genome guided Trinity across all the expression bins
and MA fragment length categories. Overall, these three assemblers outperformed Trans-ABySS and
Oases in terms of higher median MA fragment lengh recovery and tighter distribution around the
median (Figure 3).}

\bigskip
{\selectlanguage{english}\itshape
Mis-assembly statistics}

{\selectlanguage{english}
We used Megablast-based mapping to evaluate the accuracy of assembled fragments, and determine
whether each assembled fragment belonged to a single MA source or was a chimera across multiple
sources (mis-assembled). We classified the assembled fragments into three categories,
{\textgreater}=90\%, between 60-90\%, and {\textless}60\% based on the extent of their lengths
mapped to any single MA fragment. They were further classified into various assembled length
categories to check their relationship with assembly quality, before and after removing redundancy.
Trans-ABySS and Oases showed relatively higher number of mis-assembled fragments (between 60-90\%,
and {\textless}60\% mapping) in the 200-400bp and 500-600bp ranges respectively (Figure 4). In the
case of Trans-ABySS, the extent of mis-assembly decreased after removing redundancy, whereas in the
case of Oases, it went up. We did not observe any difference in the extent of mis-assembly, pre-
and post-redundancy removal, for any other assembler (Figure 4). The extent of mis-assembly was
over-estimated with Tophat1-Cufflinks as it surpassed the length of the MA fragment (Table 1). For
Trinity, SOAPdenovo-Trans and Genome guided Trinity, there was a clear trend of decreasing
mis-assembly with increasing assembled fragment length (Figure 4). The degree of mis-assembly was
lesser for SOAPdenovo-Trans than for Trinity or Genome guided Trinity for the shorter fragments. }

\bigskip

{\selectlanguage{english}\itshape
Recovery of isoforms}

{\selectlanguage{english}
Next, we compared the extent of recovery of known isoforms from the transcripts assembled with
various assemblers. We had a total of 24,846 exons in the MA fragments corresponding to 2,970
isoforms. We mapped assembled fragments from all assemblers to these exons, using Megablast, and
maximized exon coverage. Trinity, followed by Genome guided Trinity and SOAPdenovo-Trans had the
highest numbers of isoforms in the 80-100\% length recovery category (Figure 5). We observed a
correlation between the median pnc for each recovery category and the numbers of exons recovered
per isoform. The median pnc was in the range of 2-3 for unrecovered isoforms and 17-23 for those
which were fully recovered or close to fully recovered by all assemblers. Trans-ABySS and Oases,
which showed a relatively lower recovery, correlated with a higher median pnc, suggesting a higher
threshold of pnc needed for good recovery by those assemblers (Figure 5).}

\begin{figure}[htbp]
\begin{flushleft}
\includegraphics[angle=90]{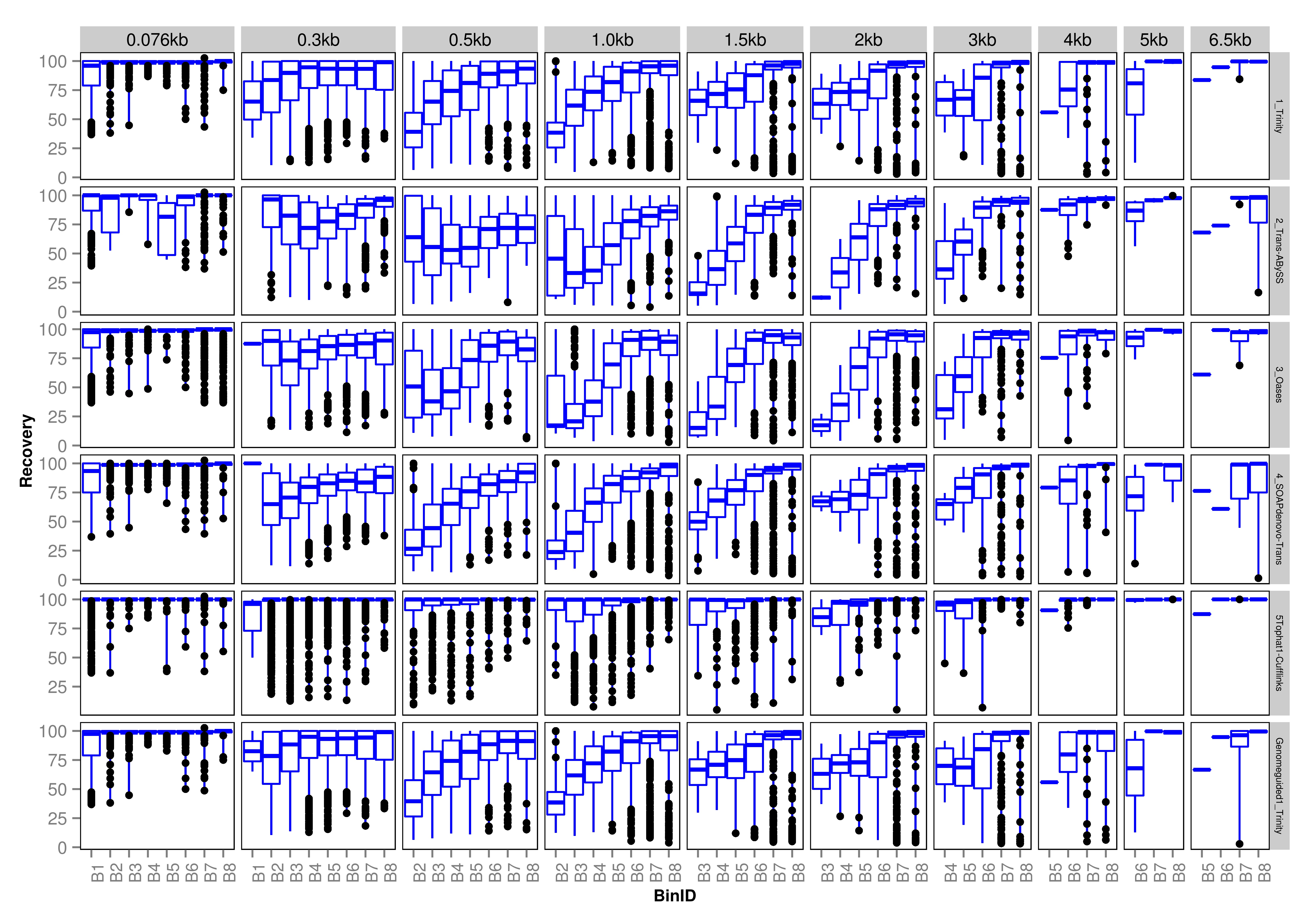}
\caption{Recovered length (\%) of MA fragments.}
\end{flushleft}
\end{figure}

\begin{figure}[htbp]
\begin{flushleft}
\includegraphics[angle=90]{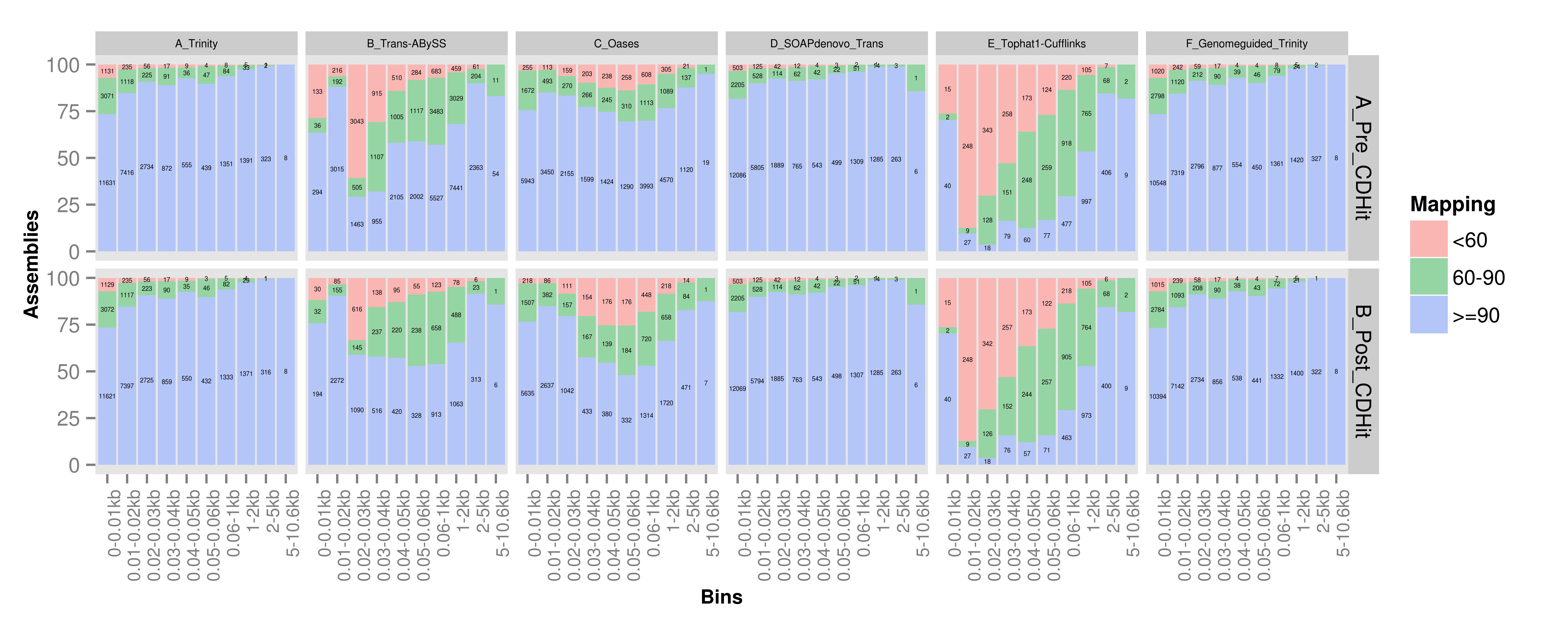}
\caption{Assembly mapping statistics.}
\end{flushleft}
\end{figure}

\begin{figure}[htbp]
\begin{flushleft}
\includegraphics[scale=0.6]{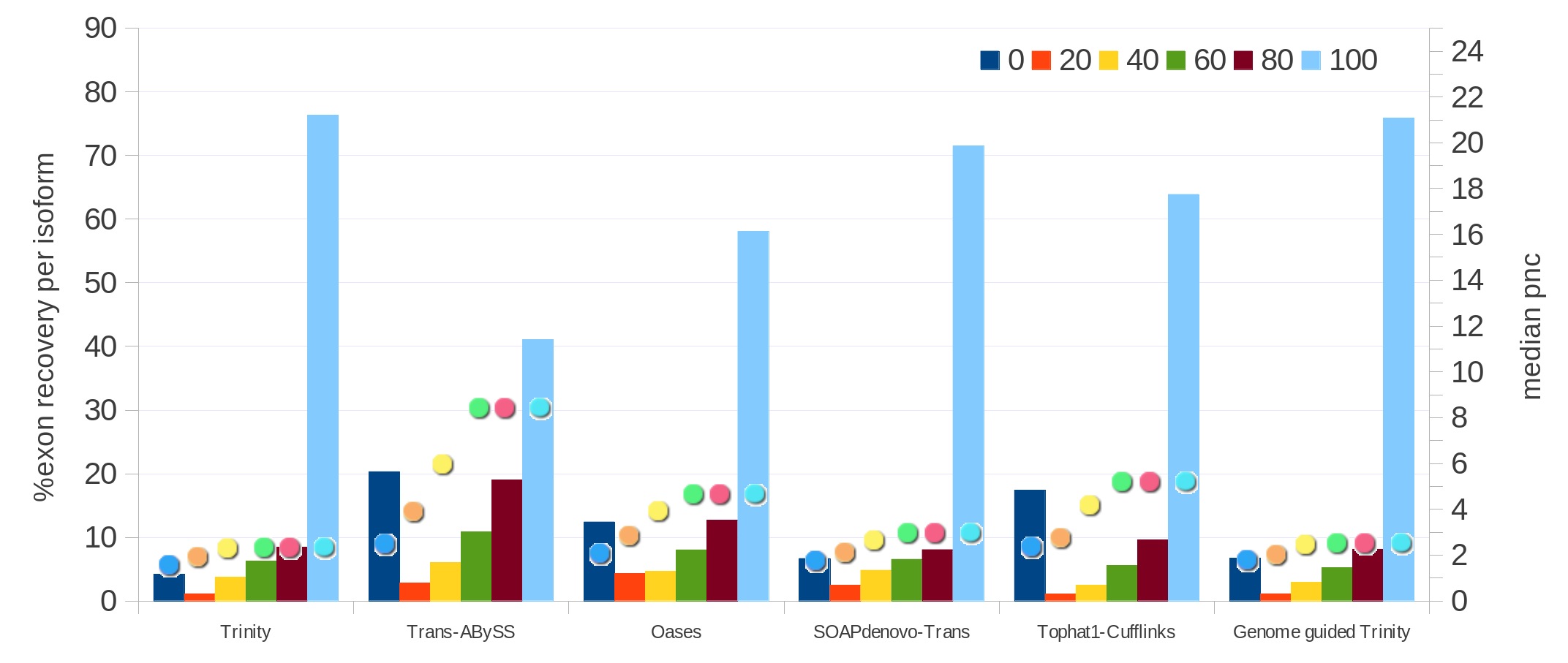}
\caption{Isoform recovery statistics.}
\end{flushleft}
\end{figure}

\bigskip

{\selectlanguage{english}
In order to test whether our understanding of isoform recovery from simulated reads holds true for a
real dataset, we measured the recovery of shared and unique assembled transcripts for the shared
and unique regions of the hox gene cluster in zebrafish (see Methods for details). We observed that
across assemblers, Trinity, followed by Genome guided Trinity and SOAPdenovo-Trans recovered most
and closer to full length transcripts (Figures 6-8). Tophat1-Cufflinks recovered mostly full-length
transcripts, but fewer in numbers. Trans-ABySS and Oases recovered fewer and truncated transcripts.
The length recovery (\%) by all assemblers except Tophat1-Cufflinks displayed a dependency on pnc
which was more obvious in the recovery of shared regions of transcripts, as expected due to a wider
range of read depth (Figures 6-7).}

\bigskip

{\selectlanguage{english}\bfseries\itshape
Augmenting transcriptome assembly }

{\selectlanguage{english}
Since each assembler produced a set of unique transcripts or fragments, we proceeded towards
augmenting \foreignlanguage{english}{the assemblies, one with another. We ruled out most
combinations of assemblers, and chose Trinity and Tophat1- Cufflinks, as Trinity produced most
number of transcript fragments and TopHat-Cufflinks the most number of full-length transcripts.
After aligning the Trinity transcript fragments to the Tophat1-Cufflinks transcripts, we obtained a
cumulative size increase of 1.37 Mbp (\~{}20\% of original Trinity assembly size), unique to
Tophat1-Cufflinks,. We observed an increase of 1,377 in number of MA fragments and 5,23,127nt in
total cumulative length recovered after augmenting Trinity transcript fragments with
Tophat1-Cufflinks-specific transcript regions. Relaxing the stringency of Megablast word size, from
28 to lower, would have increased our MA length recovery further, at the expense of losing isoform
reconstruction capability and sensitivity to variation in read-depth, maintained by the fragmented
structure of Trinity transcriptome assembly. We further observed that the median length recovery
for the augmented Trinity was overall better than Trinity by itself (Figure 9). For transcripts
longer than 1500nt, this augmentation yielded full-length transcripts across all expression levels
(Figure 9). Even for transcripts that are of sizes around 500nt, we saw the advantage of
augmentation, at least for recovering more transcripts that are towards full length. The outliers
were fewer in augmented Trinity than Tophat1-Cufflinks in all length categories. The improved
recovery with augmented }\foreignlanguage{english}{Trinity proved that an integrative approach was
useful, particularly when one has access to genome in addition to the RNA-seq reads.}}

\bigskip
\begin{figure}[htbp]
\begin{flushleft}
\includegraphics[scale=0.6]{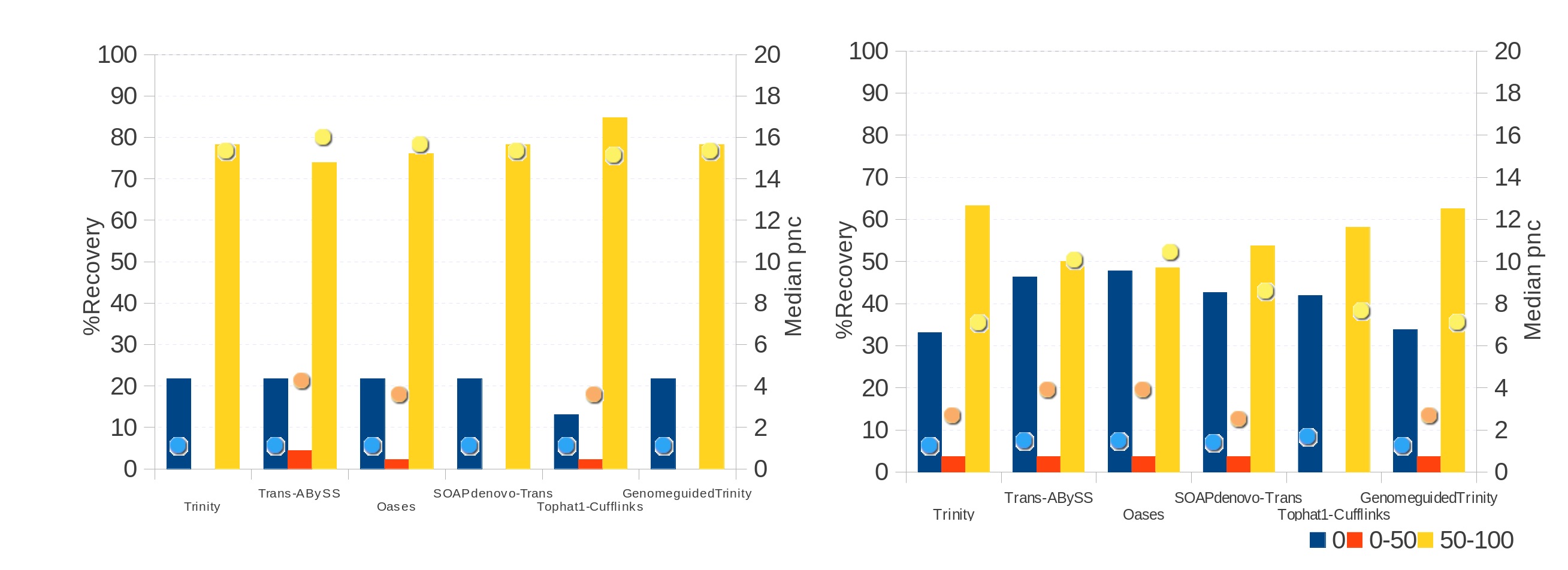}
\caption{Recovery of transcripts in the zebrafish hox gene cluster.}
\end{flushleft}
\end{figure}

\begin{figure}[htbp]
\begin{flushleft}
\includegraphics[scale=0.6]{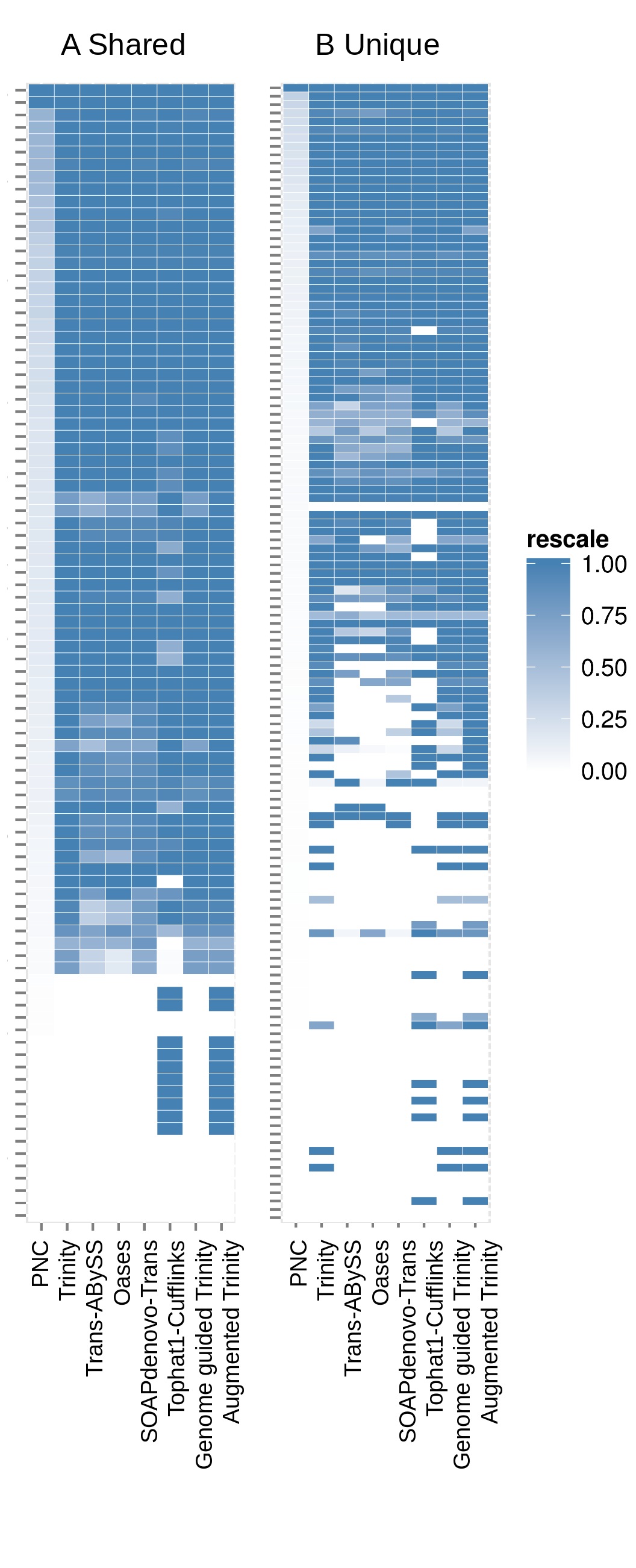}
\caption{Heatmap analyses of \% length recovery.}
\end{flushleft}
\end{figure}

\begin{figure}[htbp]
\begin{flushleft}
\includegraphics[scale=0.6]{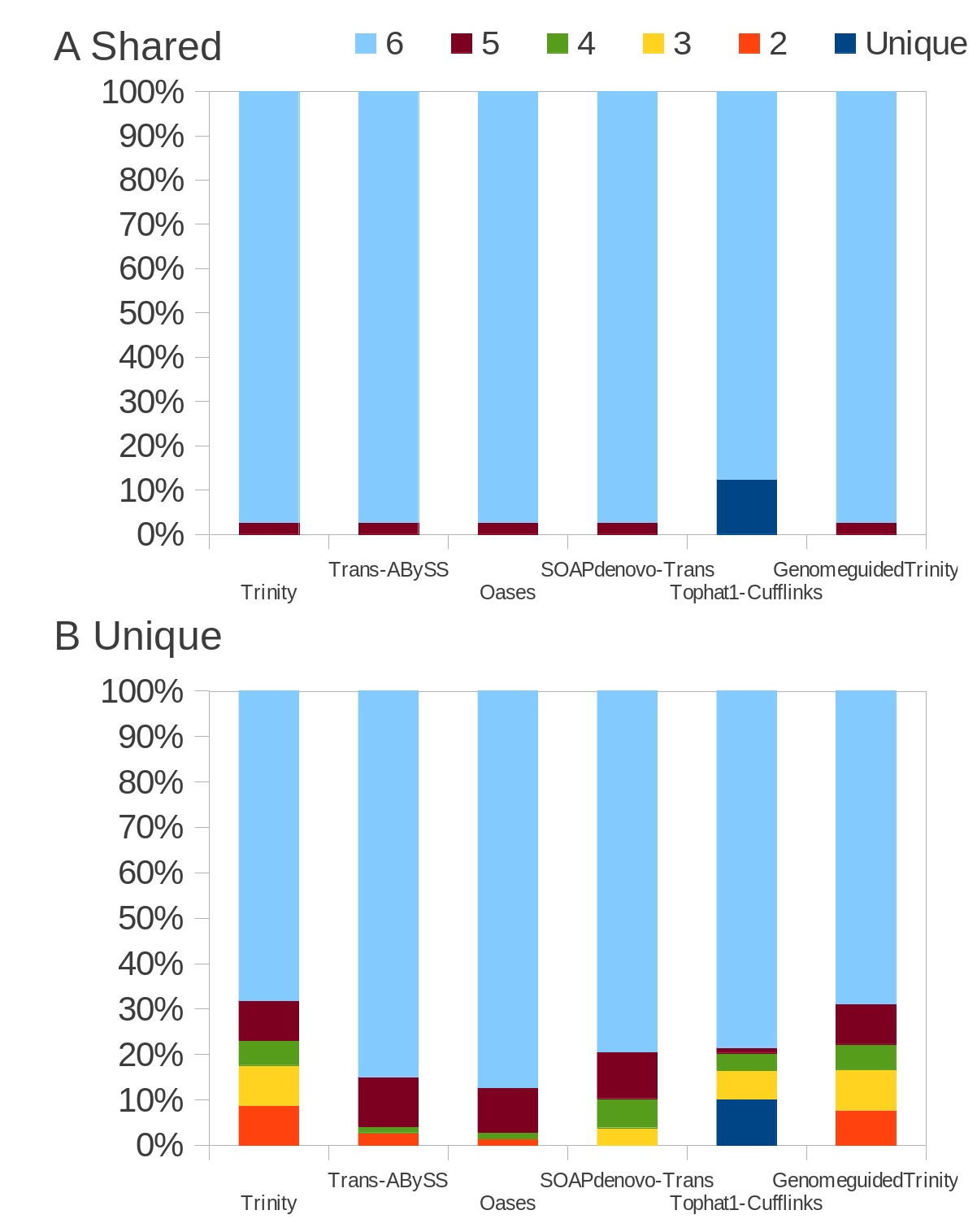}
\caption{Intersection of recovered transcripts.}
\end{flushleft}
\end{figure}

{\selectlanguage{english}\bfseries
Discussion}

{\selectlanguage{english}
RNA-seq using next-generation sequencing is a powerful technology to understand the transcriptome of
an organism. Although, genome guided assemblers like Tophat-Cufflinks can assemble full length
transcripts, most \textit{de novo} approaches and even Genome guided Trinity, where the genome is
used only to partition the RNA-seq reads, are useful in detecting novel transcripts. In our
comparative study, we found that each assembler produced a set of unique transcripts or fragments,
especially at lower levels of expression (Figure 2). Hence, we started by asking a question whether
one can obtain a better transcriptome assembly when augmenting the \textit{de novo }assembly with
that from a genome-guided approach. While this was a reasonable question to ask, the presence of
multiple tools, and hence errors associated with each of them, compounded the problem. Therefore,
we started by finding out the efficiency of individual tools, the ones that are popularly used by
the community, with the hope that we could choose from assemblers and combine the results from them
without compounding additional errors in the final assembly.}

\bigskip

{\selectlanguage{english}
A lower threshold for minimum fragment length allows one to retain valid assemblies (as demonstrated
in Figure 3) at the expense of increasing errors. However, in our observation, the errors are
comparatively lesser than the number of valid assemblies (Figure 4). The extent of mis-assembly by
different assemblers was different. Oases and Trans-ABySS resulted in more mis-assemblies than the
other tools (Figure 4). We suspected that a large number of redundant transcripts produced by Oases
and Trans-ABySS were possibly mis-assembled. Interestingly, however, when we compared the
mis-assembly statistics before and after redundancy removal, we found that Oases had more
mis-assembly in the non-redundant regions of the transcriptome than Trans-ABySS (Figure 4) pointing
towards assembly of more number of chimeric transcripts.}

\begin{figure}[htbp]
\begin{flushleft}
\includegraphics[angle=90]{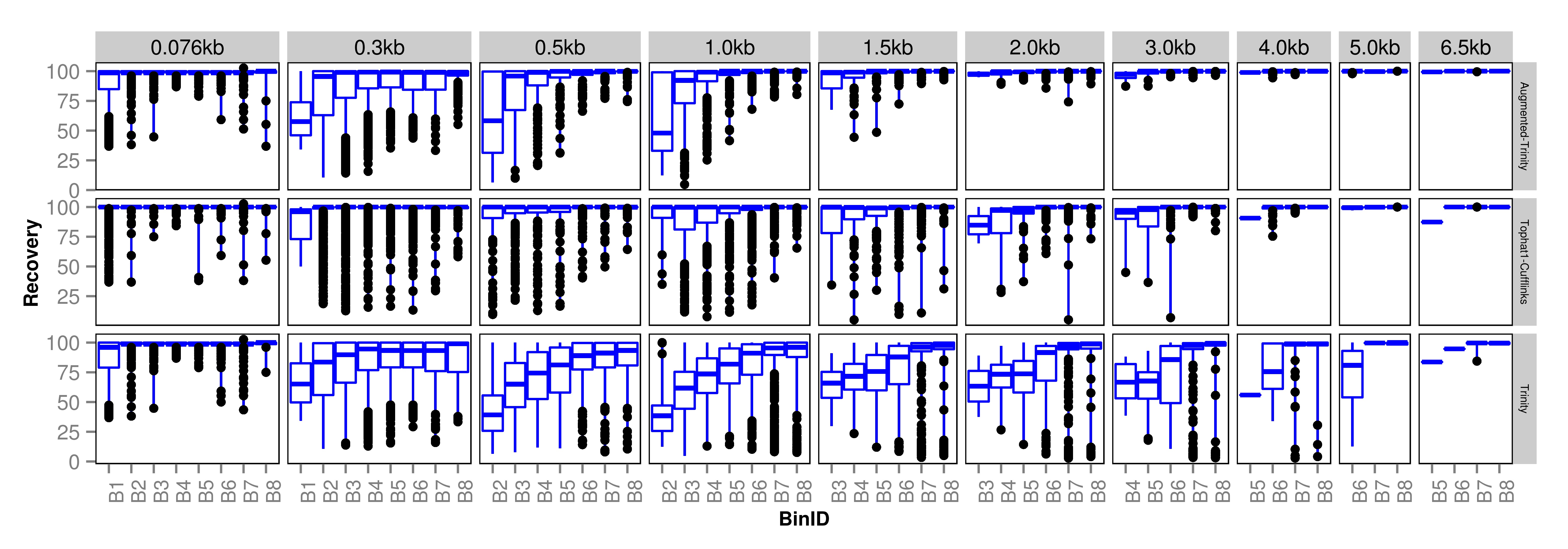}
\caption{Intersection of recovered transcripts.}
\end{flushleft}
\end{figure}
\begin{figure}[htbp]
\begin{flushleft}
\includegraphics[scale=0.8]{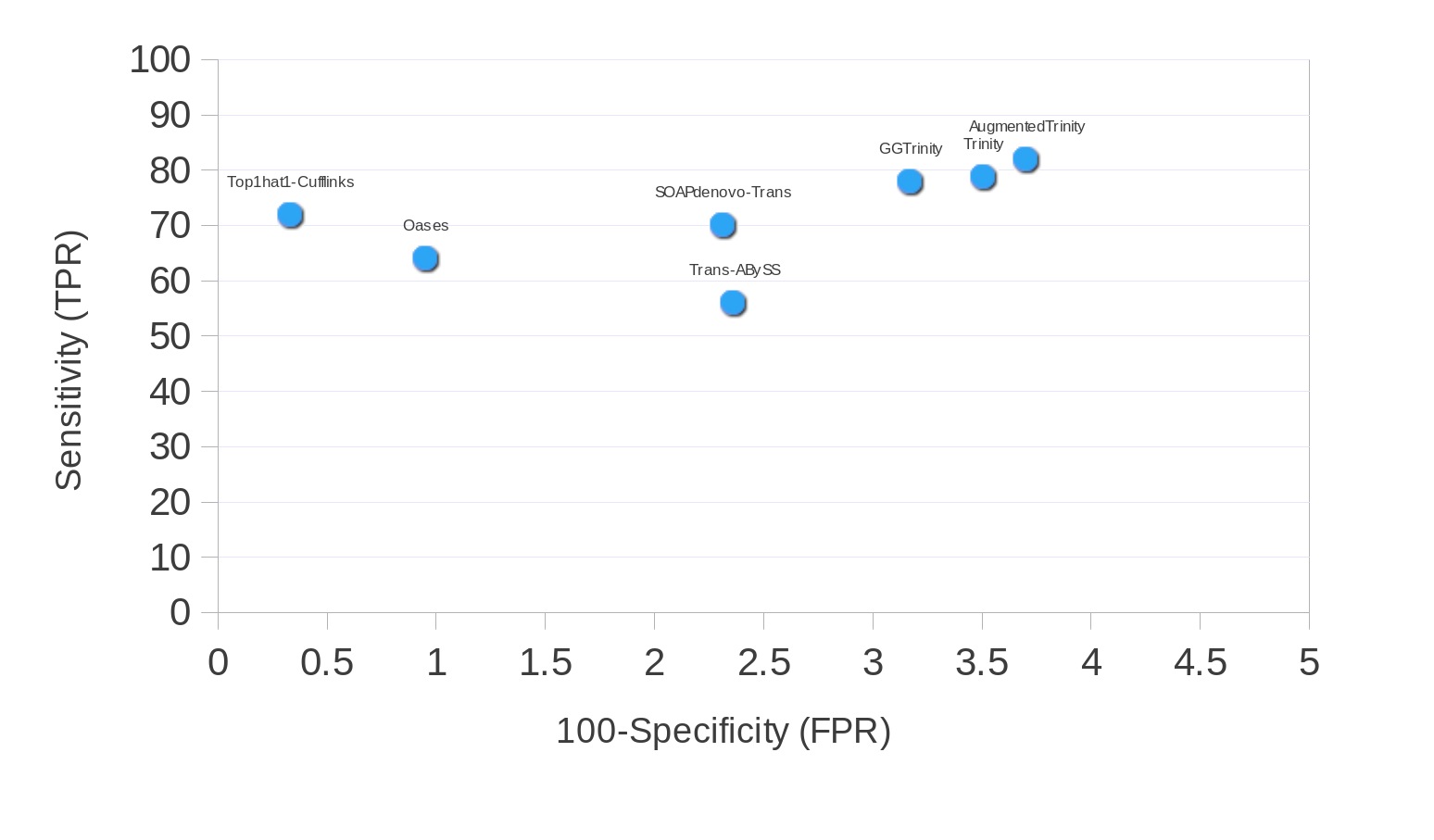}
\caption{Intersection of recovered transcripts.}
\end{flushleft}
\end{figure}
\bigskip

{\selectlanguage{english}
\foreignlanguage{english}{Due to higher sub-sequence similarity in isoforms, which contain more
shared regions than non-isoforms, the chances of mis-assembly are greater. }\ The shared regions
among transcripts also pose additional difficulty in discerning their true identity at the time of
estimating recovery using mapping. We analyzed the shared and unique regions within isoforms
separately in order to distinguish their recovery and underlying pnc patterns.
\foreignlanguage{english}{All measures of pnc used for comparisons in this paper are based on the
reads mapped to the MA as we were left with the option of remapping the reads to the assemblies in
order to calculate the expression level, i.e. pnc in the absence of read tracking during
transcriptome assembly. }Re-mapping of reads is typically done using aligners that might result in
multiple hits. This compels the user to arbitrarily assign the pnc to a transcript fragment, which
may or may not be the same as the pnc estimated by an assembler. With simulated reads, where we
knew the actual pnc of MA fragments, relationship between the actual (simulated) and estimated pnc
using Megablast could be correlated. As expected, we found this to be true (Supplementary Figure
2). Based on this, we expected that the estimated pnc of transcript fragments by any assembler to
be positively correlated with the actual pnc. In addition to the simulated dataset, we used the
zebrafish hox gene cluster for transcript recovery analysis. Vertebrate Hox
\foreignlanguage{english}{genes are known to be involved during development, are arranged in sets
of uninterrupted clusters, and are in most cases expressed in a collinear fashion \ making it an
ideal gene cluster candidate for our assembly analysis. We used RNA-seq reads from zebrafish to
perform the assembly and derived assembled fragments related to the hox gene cluster to perform
recovery analysis. We expected the s}hared regions had sequencing reads at a greater read depth
than the unique regions. Indeed, we found this to be reflected in terms of a greater median pnc for
the shared regions in the {\textgreater}50\% recovery category (Figures 6 and 7). }

\bigskip

{\selectlanguage{english}
\foreignlanguage{english}{N50 is an assembly attribute widely used to compare the quality of a
transcriptome assembly. In the absence of the knowledge of actual length distribution in the
sequenced dataset (transcripts sequenced), it is generally assumed that a higher N50 is correlated
with a better assembly length-recovery/contiguity. }We know that errors and redundancy in the
transcriptome assembly affect its total size. In addition, the lengths of assembly fragments were
also variable (e.g. Tophat1-Cufflinks produced more full length transcripts and Trinity produced
more number of small fragments than the rest of the assemblers). Therefore, basing a quality metric
on the total assembly size to benchmark the assembler, like what the N50 does, can be highly
inaccurate. Instead, basing the quality metric on the expected size of the transcriptome (NT50) or
regions of the transcriptome that is covered by sequencing reads, as N\textsubscript{(MA)}50
suggests in simulated reads, tend to provide a more accurate benchmark. Indeed, we found that the
N\textsubscript{(MA)}50 to accurately reflect the recovery quotient of an assembler (Table 1). The
reduction in N\textsubscript{(MA)}50 for well-performing assemblers (Trinity, Genome guided Trinity
and SOAPdenovo-Trans) when compared to that for the MA, was in proportion to the relative reduction
in their assembly numbers in comparison to the number of MA fragments (Table 1). In addition to
N\textsubscript{(MA)}50, we found that the median, inter-quartile range, minimum, maximum and
outliers for transcript assembly length were more useful in describing a transcriptome assembly. }

\bigskip

{\selectlanguage{english}
\foreignlanguage{english}{Mis-assembly can occur as a result of sub-sequence similarity within
reads, which manifests as highly branched nodes in a de Bruijn graph. This subsequence similarity,
along with mismatches/errors in sequencing reads, can also cause spurious blast hits that are seen
as outliers observed in the box and whisker plot for MA recovery, represented in Figure 3. Spurious
blast hits due to the former tend to span from Megablast word size of 28bp onwards. }We observed
fewer outliers of the first kind in multi-kmer based approaches like Oases and Trans-ABySS.
SOAPdenovo-Trans also appears to contain fewer spurious blast hits, especially for the 76bp MA
length category. It is known that SOAPdenovo algorithm chooses to discard a repetitive node if
there is unequal read support on edges to and from that node, and only build parallel paths
carrying the node if there are equal number of reads on either edges . Based on our observations of
low redundancy in mapped assemblies in the 76bp MA length category (Figure 3) and higher \% of
assembly mapping correctly to a single MA fragment in the lower assembled fragment length
categories (Figure 4), it is possible that SOAPdenovo-Trans discarded repetitive sequences from the
assembly for these length categories. TransABySS, prior to eliminating redundancy, showed a higher
frequency for assembled fragments lengths {\textgreater} 200bp (Figure 1A). The frequency of
assembled fragments in this length range was much lower, post CD-HIT-EST (Figure 1B), suggesting
the sequence redundancy in this range. Hence, we believe that there was relationship between
redundancy and recovery of assembled fragments in our study. }
{\selectlanguage{english}
In our study, we found Trinity to perform best in transcript recovery across all expression levels
(Figures 2 and 3). The median length recovery by Tophat1-Cufflinks was always \~{}100\% (Figure 3).
We chose to combine the Trinity assemblies with that from the Tophat1-Cufflinks assembly pipeline. This resulted in
1300 more transcript fragements corresponding to an cumulative increase of nearly 0.5million nt in
length. The
process resulted in an augmented assembly with even greater sensitivity and only minimally
compromising its specificity, while maintaining the expression-based fragmented assembly structure of
Trinity post- augmentation (Figure 10). The improved recovery post-augmentation proved that an integrative
approach may be employed in recovering more transcript fragments, particularly when one has access
to genome assembly in addition to RNA-seq reads. }

\bigskip

{\selectlanguage{english}\bfseries
Acknowledgement}

{\selectlanguage{english}
The authors would like to thank Professor N. Yathindra for encouragement.}

\bigskip

{\selectlanguage{english}\bfseries
Funding}

{\selectlanguage{english}
Research is funded by Department of Electronics and Information Technology, Government of India (Ref
No:18(4)/2010-E-Infra., 31-03-2010) \ and Department of IT, BT and ST, Government of Karnataka,
India (Ref No:3451-00-090-2-22) under the {\textquotedblleft}Bio-IT Project{\textquotedblright}.}

\bigskip

{\selectlanguage{english}\bfseries
Author Contributions}

{\selectlanguage{english}
Conceived the project: BP. Designed and analyzed the data: PJ and NMK. Wrote the paper: PJ, NMK \&
BP.}

\bigskip

{\selectlanguage{english}\bfseries
References}

{\selectlanguage{english}
\foreignlanguage{english}{Altschul SF, Gish W, Miller W, Myers EW, and Lipman DJ. 1990. Basic local
alignment search tool.}\foreignlanguage{english}{\textit{ J Mol Biol}}\foreignlanguage{english}{
215:403-410.}}

{\selectlanguage{english}
\foreignlanguage{english}{Birol I, Jackman SD, Nielsen CB, Qian JQ, Varhol R, Stazyk G, Morin RD,
Zhao Y, Hirst M, Schein JE et al. . 2009. De novo transcriptome assembly with
ABySS.}\foreignlanguage{english}{\textit{ Bioinformatics}}\foreignlanguage{english}{
25:2872-2877.}}

{\selectlanguage{english}
\foreignlanguage{english}{Collins LJ, Biggs PJ, Voelckel C, and Joly S. 2008. An approach to
transcriptome analysis of non-model organisms using short-read
sequences.}\foreignlanguage{english}{\textit{ Genome Inform}}\foreignlanguage{english}{ 21:3-14.}}

{\selectlanguage{english}
\foreignlanguage{english}{Corredor-Adamez M, Welten MC, Spaink HP, Jeffery JE, Schoon RT, de Bakker
MA, Bagowski CP, Meijer AH, Verbeek FJ, and Richardson MK. 2005. Genomic annotation and
transcriptome analysis of the zebrafish (Danio rerio) hox complex with description of a novel
member, hox b 13a.}\foreignlanguage{english}{\textit{ Evol Dev}}\foreignlanguage{english}{
7:362-375.}}

{\selectlanguage{english}
\foreignlanguage{english}{DeRisi J, Penland L, Brown PO, Bittner ML, Meltzer PS, Ray M, Chen Y, Su
YA, and Trent JM. 1996. Use of a cDNA microarray to analyse gene expression patterns in human
cancer.}\foreignlanguage{english}{\textit{ Nat Genet}}\foreignlanguage{english}{ 14:457-460.}}

{\selectlanguage{english}
\foreignlanguage{english}{Egan AN, Schlueter J, and Spooner DM. 2012. Applications of
next-generation sequencing in plant biology.}\foreignlanguage{english}{\textit{ Am J
Bot}}\foreignlanguage{english}{ 99:175-185.}}

{\selectlanguage{english}
\foreignlanguage{english}{Gibbons JG, Janson EM, Hittinger CT, Johnston M, Abbot P, and Rokas A.
2009. Benchmarking next-generation transcriptome sequencing for functional and evolutionary
genomics.}\foreignlanguage{english}{\textit{ Mol Biol Evol}}\foreignlanguage{english}{
26:2731-2744.}}

{\selectlanguage{english}
\foreignlanguage{english}{Golub TR, Slonim DK, Tamayo P, Huard C, Gaasenbeek M, Mesirov JP, Coller
H, Loh ML, Downing JR, Caligiuri MA et al. . 1999. Molecular classification of cancer: class
discovery and class prediction by gene expression monitoring.}\foreignlanguage{english}{\textit{
Science}}\foreignlanguage{english}{ 286:531-537.}}

{\selectlanguage{english}
\foreignlanguage{english}{Grabherr MG, Haas BJ, Yassour M, Levin JZ, Thompson DA, Amit I, Adiconis
X, Fan L, Raychowdhury R, Zeng Q et al. . 2011. Full-length transcriptome assembly from RNA-Seq
data without a reference genome.}\foreignlanguage{english}{\textit{ Nat
Biotechnol}}\foreignlanguage{english}{ 29:644-652.}}

{\selectlanguage{english}
\foreignlanguage{english}{Griebel T, Zacher B, Ribeca P, Raineri E, Lacroix V, Guigo R, and Sammeth
M. 2012. Modelling and simulating generic RNA-Seq experiments with the flux
simulator.}\foreignlanguage{english}{\textit{ Nucleic Acids Res}}\foreignlanguage{english}{
40:10073-10083.}}

{\selectlanguage{english}
\foreignlanguage{english}{Gruenheit N, Deusch O, Esser C, Becker M, Voelckel C, and Lockhart P.
2012. Cutoffs and k-mers: implications from a transcriptome study in allopolyploid
plants.}\foreignlanguage{english}{\textit{ BMC Genomics}}\foreignlanguage{english}{ 13:92.}}

{\selectlanguage{english}
\foreignlanguage{english}{Kim D, Pertea G, Trapnell C, Pimentel H, Kelley R, and Salzberg SL. 2013.
TopHat2: accurate alignment of transcriptomes in the presence of insertions, deletions and gene
fusions.}\foreignlanguage{english}{\textit{ Genome Biol}}\foreignlanguage{english}{ 14:R36.}}

{\selectlanguage{english}
\foreignlanguage{english}{Kuraku S, and Meyer A. 2009. The evolution and maintenance of Hox gene
clusters in vertebrates and the teleost-specific genome
duplication.}\foreignlanguage{english}{\textit{ Int J Dev Biol}}\foreignlanguage{english}{
53:765-773.}}

{\selectlanguage{english}
\foreignlanguage{english}{Langmead B. 2010. Aligning short sequencing reads with
Bowtie.}\foreignlanguage{english}{\textit{ Curr Protoc Bioinformatics}}\foreignlanguage{english}{
Chapter 11:Unit 11 17.}}

{\selectlanguage{english}
\foreignlanguage{english}{Li R, Yu C, Li Y, Lam TW, Yiu SM, Kristiansen K, and Wang J. 2009. SOAP2:
an improved ultrafast tool for short read alignment.}\foreignlanguage{english}{\textit{
Bioinformatics}}\foreignlanguage{english}{ 25:1966-1967.}}

{\selectlanguage{english}
\foreignlanguage{english}{Li R, Zhu H, Ruan J, Qian W, Fang X, Shi Z, Li Y, Li S, Shan G,
Kristiansen K et al. . 2010. De novo assembly of human genomes with massively parallel short read
sequencing.}\foreignlanguage{english}{\textit{ Genome Res}}\foreignlanguage{english}{ 20:265-272.}}

{\selectlanguage{english}
\foreignlanguage{english}{Li W, and Godzik A. 2002. Discovering new genes with advanced homology
detection.}\foreignlanguage{english}{\textit{ Trends Biotechnol}}\foreignlanguage{english}{
20:315-316.}}

{\selectlanguage{english}
\foreignlanguage{english}{Martin JA, and Wang Z. 2011. Next-generation transcriptome
assembly.}\foreignlanguage{english}{\textit{ Nat Rev Genet}}\foreignlanguage{english}{
12:671-682.}}

{\selectlanguage{english}
\foreignlanguage{english}{Mitchell W. 2011. Natural products from synthetic
biology.}\foreignlanguage{english}{\textit{ Curr Opin Chem Biol}}\foreignlanguage{english}{
15:505-515.}}

{\selectlanguage{english}
\foreignlanguage{english}{Mortazavi A, Williams BA, McCue K, Schaeffer L, and Wold B. 2008. Mapping
and quantifying mammalian transcriptomes by RNA-Seq.}\foreignlanguage{english}{\textit{ Nat
Methods}}\foreignlanguage{english}{ 5:621-628.}}

{\selectlanguage{english}
\foreignlanguage{english}{Mundry M, Bornberg-Bauer E, Sammeth M, and Feulner PG. 2012. Evaluating
characteristics of de novo assembly software on 454 transcriptome data: a simulation
approach.}\foreignlanguage{english}{\textit{ PLoS One}}\foreignlanguage{english}{ 7:e31410.}}

{\selectlanguage{english}
\foreignlanguage{english}{Nagarajan N, and Pop M. 2013. Sequence assembly
demystified.}\foreignlanguage{english}{\textit{ Nat Rev Genet}}\foreignlanguage{english}{
14:157-167.}}

{\selectlanguage{english}
\foreignlanguage{english}{Ozsolak F, and Milos PM. 2011. RNA sequencing: advances, challenges and
opportunities.}\foreignlanguage{english}{\textit{ Nat Rev Genet}}\foreignlanguage{english}{
12:87-98.}}

{\selectlanguage{english}
\foreignlanguage{english}{Roberts A, Pimentel H, Trapnell C, and Pachter L. 2011. Identification of
novel transcripts in annotated genomes using RNA-Seq.}\foreignlanguage{english}{\textit{
Bioinformatics}}\foreignlanguage{english}{ 27:2325-2329.}}

{\selectlanguage{english}
\foreignlanguage{english}{Robertson G, Schein J, Chiu R, Corbett R, Field M, Jackman SD, Mungall K,
Lee S, Okada HM, Qian JQ et al. . 2010. De novo assembly and }\foreignlanguage{english}{analysis of
RNA-seq data.}\foreignlanguage{english}{\textit{ Nat Methods}}\foreignlanguage{english}{
}\foreignlanguage{english}{7:909-912.}}

{\selectlanguage{english}
\foreignlanguage{english}{Salzberg SL, Sommer DD, Puiu D, and Lee VT. 2008. Gene-boosted assembly of
a novel bacterial genome from very short reads.}\foreignlanguage{english}{\textit{ PLoS Comput
Biol}}\foreignlanguage{english}{ 4:e1000186.}}

{\selectlanguage{english}
\foreignlanguage{english}{Schena M, Shalon D, Davis RW, and Brown PO. 1995. Quantitative monitoring
of gene expression patterns with a complementary DNA microarray.}\foreignlanguage{english}{\textit{
Science}}\foreignlanguage{english}{ 270:467-470.}}

{\selectlanguage{english}
\foreignlanguage{english}{Schena M, Shalon D, Heller R, Chai A, Brown PO, and Davis RW. 1996.
Parallel human genome analysis: microarray-based expression monitoring of 1000
genes.}\foreignlanguage{english}{\textit{ Proc Natl Acad Sci U S A}}\foreignlanguage{english}{
93:10614-10619.}}

{\selectlanguage{english}
\foreignlanguage{english}{Schulz MH, Zerbino DR, Vingron M, and Birney E. 2012. Oases: robust de
novo RNA-seq assembly across the }\foreignlanguage{english}{dynamic range of expression
levels.}\foreignlanguage{english}{\textit{ Bioinformatics}}\foreignlanguage{english}{
28:1086-1092.}}

{\selectlanguage{english}
\foreignlanguage{english}{Shendure J, and Lieberman Aiden E. 2012. The expanding scope of DNA
sequencing.}\foreignlanguage{english}{\textit{ Nat Biotechnol}}\foreignlanguage{english}{
30:1084-1094.}}

{\selectlanguage{english}
\foreignlanguage{english}{Simon SA, Zhai J, Nandety RS, McCormick KP, Zeng J, Mejia D, and Meyers
BC. 2009. Short-read sequencing technologies for transcriptional
analyses.}\foreignlanguage{english}{\textit{ Annu Rev Plant Biol}}\foreignlanguage{english}{
60:305-333.}}

{\selectlanguage{english}
\foreignlanguage{english}{Simpson JT, Wong K, Jackman SD, Schein JE, Jones SJ, and Birol I. 2009.
ABySS: a parallel assembler for short read sequence data.}\foreignlanguage{english}{\textit{ Genome
Res}}\foreignlanguage{english}{ 19:1117-1123.}}

{\selectlanguage{english}
\foreignlanguage{english}{Toth AL, Varala K, Newman TC, Miguez FE, Hutchison SK, Willoughby DA,
Simons JF, Egholm M, Hunt JH, Hudson ME et al. . 2007. Wasp gene expression supports an
evolutionary link between maternal behavior and eusociality.}\foreignlanguage{english}{\textit{
Science}}\foreignlanguage{english}{ 318:441-444.}}

{\selectlanguage{english}
\foreignlanguage{english}{Trapnell C, Pachter L, and Salzberg SL. 2009. TopHat: discovering splice
junctions with RNA-Seq.}\foreignlanguage{english}{\textit{
Bioinformatics}}\foreignlanguage{english}{ 25:1105-1111.}}

{\selectlanguage{english}
\foreignlanguage{english}{Trapnell C, Roberts A, Goff L, Pertea G, Kim D, Kelley DR, Pimentel H,
Salzberg SL, Rinn JL, and Pachter L. 2012. Differential gene and transcript expression analysis of
RNA-seq experiments with TopHat and Cufflinks.}\foreignlanguage{english}{\textit{ Nat
Protoc}}\foreignlanguage{english}{ 7:562-578.}}

{\selectlanguage{english}
\foreignlanguage{english}{Trapnell C, Williams BA, Pertea G, Mortazavi A, Kwan G, van Baren MJ,
Salzberg SL, Wold BJ, and Pachter L. 2010. Transcript assembly and quantification by RNA-Seq
reveals unannotated transcripts and isoform switching during cell
differentiation.}\foreignlanguage{english}{\textit{ Nat Biotechnol}}\foreignlanguage{english}{
28:511-515.}}

{\selectlanguage{english}
\foreignlanguage{english}{TrinityTeam. 2013. Genome-guided Trinity.}\foreignlanguage{english}{
http://trinityrnaseq.sourceforge.net/genome\_guided\_trinity.html}}

{\selectlanguage{english}
\foreignlanguage{english}{Waern K, Nagalakshmi U, and Snyder M. 2011. RNA
sequencing.}\foreignlanguage{english}{\textit{ Methods Mol Biol}}\foreignlanguage{english}{
759:125-132.}}

{\selectlanguage{english}
\foreignlanguage{english}{Wang Z, Gerstein M, and Snyder M. 2009. RNA-Seq: a revolutionary tool for
transcriptomics.}\foreignlanguage{english}{\textit{ Nat Rev Genet}}\foreignlanguage{english}{
10:57-63.}}

{\selectlanguage{english}
\foreignlanguage{english}{Wu TD, and Nacu S. 2010. Fast and SNP-tolerant detection of complex
variants and splicing in short reads.}\foreignlanguage{english}{\textit{
Bioinformatics}}\foreignlanguage{english}{ 26:873-881.}}

{\selectlanguage{english}
\foreignlanguage{english}{Zerbino DR, and Birney E. 2008. Velvet: algorithms for de novo short read
assembly using de Bruijn graphs.}\foreignlanguage{english}{\textit{ Genome
Res}}\foreignlanguage{english}{ 18:821-829.}}

\clearpage{\selectlanguage{english}\bfseries
Figure Legends}

\bigskip

{\selectlanguage{english}
\textbf{Figure 1:} Frequency distribution of lengths (bp) of MA \& assembled transcript fragments
before (A) and after (B) CDHit-EST.}

\bigskip

{\selectlanguage{english}
\textbf{Figure 2: }The number of recovered MA fragments in different expression bins. }

{\selectlanguage{english}
The recovery of MA fragments by six assemblers was estimated across eight expression level
categories (B1 - B8) binned by per-nucleotide-coverage (pnc). The recovered fragments are presented
as unique to an assembler or as an overlap between 2, 3, 4, or 5, or all the six assemblers for
each expression bin.}

\bigskip

{\selectlanguage{english}
\textbf{Figure 3:} Recovered length (\%) of MA fragments. }

{\selectlanguage{english}
MA fragments were binned into different length categories ({\textless}=76bp, 76-300bp, 300-500bp,
500bp-1kb, 1-1.5kb, 1.5-2kb, 2-3kb, 3-4kb, 4-5kb, 5-6.5kb), and the length recovered for each
assembler across expression bins (B1-B8) was visualized separately for each MA length bin. The
black dots represent the outliers, the boxes represent the 25\% (Q1)-75\% (Q3) interquartile range
(IQR), the middle line in the boxes represents the median, and the blue solid lines represent the
whiskers from these boxes till}

{\selectlanguage{english}
the minimum/maximum of the length range. Outliers fall below Q1 -- 1.5Å\~{}IQR or above Q3 +
1.5Å\~{}IQR. See Supplementary Methods for script used to generate the plot.}

\bigskip

{\selectlanguage{english}
\textbf{Figure 4: }Assembly mapping statistics}

{\selectlanguage{english}
The assembled fragments were classified as {\textgreater}90\%, between 60-90\% and {\textless}60\%
categories based on the fraction mapped to a single MA fragment, before (A) and after (B) removal
of redundancy.}

\bigskip

{\selectlanguage{english}
\textbf{Figure 5:} Isoform recovery statistics.}

{\selectlanguage{english}
Recovery was estimated for the number of exons per isoform in different (0\%, {\textgreater}0- 20\%,
{\textgreater}20-40\%, {\textgreater}40-60\% ,{\textgreater}60-80\% and {\textgreater}80-100\%)
recovery categories. The median pnc of isoforms for each of these categories was also estimated for
each assembler.}

\bigskip

{\selectlanguage{english}
\textbf{Figure 6:} Recovery of transcripts in the unique and shared regions of hox gene cluster in
zebrafish.}

{\selectlanguage{english}
The unique and shared regions of hox gene cluster from zebrafish Danio rerio with a minimum
per-nucleotide coverage (pnc) of 1 were extracted. The recovery was measured for these regions in
the 0\%, 0-50\% and 50-100\% recovery categories.}

\bigskip

{\selectlanguage{english}
\textbf{Figure 7:} Heatmap analyses of \%length recovery of shared and unique transcript regions of
the zebrafish hox gene cluster.}

{\selectlanguage{english}
The transcript regions were ranked in a descending order of their PNCs.}

\bigskip

{\selectlanguage{english}
\textbf{Figure 8: }Intersection histogram of recovered transcripts from the shared and unique
regions of the zebrafish hox gene cluster}

\bigskip

{\selectlanguage{english}
\textbf{Figure 9:} Recovered length (\%) of model assembly (MA) fragments after augmenting with
Trinity-assembled fragments.}

{\selectlanguage{english}
Mapping of Trinity assembled transcript fragments with the transcripts assembled by
Tophat1-Cufflinks. The transcript regions unique to Tophat1-Cufflinks were used to augment the
Trinity assembly. The length recovery was visualized in the same}

{\selectlanguage{english}
manner as described in Figure 3.}

\bigskip

{\selectlanguage{english}
\textbf{Figure 10: }ROC curve for transcriptome assemblers.}

{\selectlanguage{english}
The sensitvity (TPR) was estimated as \% total length recovered for each assembler out of the total
MA size. The FPR (100-specificity) was estimated as the \% assembled fragments that did not map to
the MA fragments. For Tophat1-Cufflinks assembler, and the unique regions to Tophat1-Cufflinks used
to augment Trinity, the mapping was performed with the TAIR10 transcripts.}
\newpage 
\begin{figure}[htbp]
\begin{flushleft}
\includegraphics[scale=1]{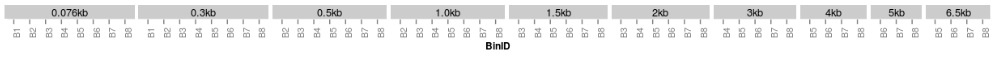}
\setcounter{figure}{0}
\makeatletter 
\renewcommand{\thefigure}{S\@arabic\c@figure}
\caption{Expression level bin assignment for MA fragments with varying length categories.}
\end{flushleft}
\end{figure}

\begin{figure}[htbp]
\begin{flushleft}
\includegraphics[scale=0.5]{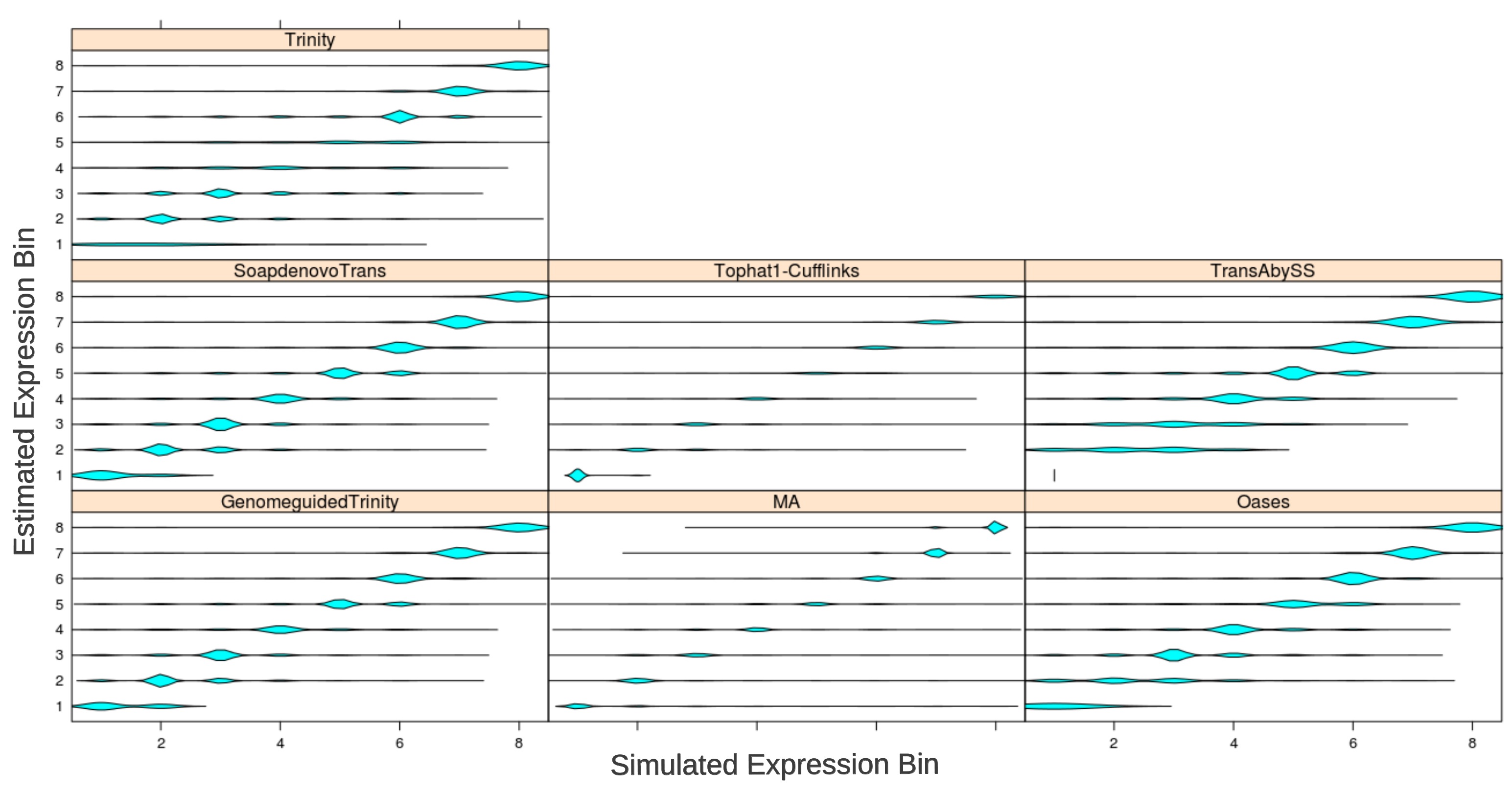}
\makeatletter 
\renewcommand{\thefigure}{S\@arabic\c@figure}
\caption{Correlation between simulated and estimated expression level binning.}
\end{flushleft}
\end{figure}

\end{document}